\documentclass[english, aps, prd, superscriptaddress,
preprintnumbers,nofootinbib,twocolumn,
]{revtex4-1}
\usepackage[T1]{fontenc}
\usepackage[utf8]{inputenc}
\usepackage{verbatim}
\usepackage{empheq}
\usepackage{amssymb}
\usepackage{amsmath}
\usepackage{amsfonts}
\usepackage{float}

\usepackage[colorlinks]{hyperref}
\hypersetup{citecolor=blue,linkcolor=blue,urlcolor=blue}

\usepackage{dcolumn}
\usepackage{graphicx}
\usepackage{etoolbox}
\usepackage{booktabs}
\usepackage{babel}

\def\ts{\textsc}

\newcommand{\LNP}{ {\Lambda} }

\newcommand{\eqspace}{\hphantom{{}={}}}

\newcommand{\chbox}{C_{\varphi\square}}
\newcommand{\cht}{C_{t\varphi}}
\newcommand{\ch}{C_{\varphi}}
\newcommand{\chg}{C_{\varphi G}}
\newcommand{\chd}{C_{\varphi D}}
\newcommand{\cg}{{C_{G}}}
\newcommand{\cll}{C_{ll}}
\newcommand{\chlthree}{{C_{\varphi l}^{(3)}}}
\newcommand{\cug}{{C_{tG}}}
\newcommand{\vev}{v_T}
\newcommand{\sqtwoGf}{{\big({\sqrt{2}G_F}\big)}{}}

\newcommand{\vpjt}{\mbox{${\varphi^\dag i\,\raisebox{2mm}{\boldmath ${}^\leftrightarrow$}\hspace{-4mm} D_\mu^{\,a}\,\varphi}$}}



\begin{document} 

\global\long\def\order#1{\mathcal{O}{\left(#1\right)}}
\global\long\def\d{\mathrm{d}}
\global\long\def\P{P}
\global\long\def\amp{{\mathcal M}}

\def\BNL{High Energy Theory Group, Department of Physics, Brookhaven 
	National Laboratory, Upton, NY 11973, USA }

\title{Double insertions of SMEFT operators in gluon fusion Higgs boson production} 

\author{Konstantin~Asteriadis}
\email[Electronic address: ]{kasteriad@bnl.gov}
\affiliation{\BNL}

\author{Sally~Dawson}
\email[Electronic address: ]{dawson@bnl.gov}
\affiliation{\BNL}

\author{Duarte~Fontes}
\email[Electronic address: ]{dfontes@bnl.gov}
\affiliation{\BNL}

\begin{abstract}
Deviations from the Standard Model (SM) can be parameterized in terms of the SM effective field theory (SMEFT), which is typically truncated at dimension-6.
Including higher dimension operators --- as well as considering simultaneous insertions of multiple dimension-6 operators --- may be necessary in some processes, in order to correctly capture the properties of the underlying UV theory.
As a step towards clarifying this in the Higgs boson production in gluon fusion process, we study double insertions of dimension-6 operators in the 1-loop virtual amplitude.
We present needed Feynman rules up to $\order{1/\LNP^4}$ and we numerically study the impact of various approximations to the $\order{1/\LNP^4}$ expansion. 
\end{abstract}

\maketitle



\section{Introduction}
\label{sec:introduction}

Current measurements of LHC experiments are in excellent agreement with theoretical predictions, but with uncertainties at the ${\cal{O}}(5-20\,\%)$ level~\cite{ATLAS:2022vkf,*CMS:2022dwd}.
As a result,  the High Luminosity LHC program will be focussed on high precision measurements.
It is expected that the experimental uncertainties will reduced to
$\order{1\,\%}$ for many observables~\cite{deBlas:2019rxi}.
 This requires precise theoretical Standard Model (SM) predictions, but also
 precise computations in specific Beyond the Standard Model (BSM) scenarios to describe potentially emerging small non-SM signatures. A more general approach is also possible; BSM physics which contains no new light particles and which respects the SM
gauge symmetries can be parameterized using the Standard Model effective field theory (SMEFT)~\cite{Brivio:2017vri}. This consists of an expansion around the SM Lagrangian $\mathcal{L}_{\text{SM}}$ in terms of an infinite tower of higher dimension operators,
\begin{align}
	\label{eq:intro:SMEFT-Lag}
	\mathcal{L} = \mathcal{L}_{\text{SM}} + \sum_{d = 5}^\infty \sum_{i} {C_i^d  O_i^d\over \LNP^{d-4}}  \, ,
\end{align}
where $\LNP$ is chosen to be the scale of new physics, $ O_i^d$ are operators of dimension $d$, and $C_i^d$ the corresponding dimensionless SMEFT Wilson coefficients (WC).
Fits to the latter have been made using Higgs, di-boson, electroweak precision, and top data~\cite{Ethier:2021bye,Ellis:2020unq,DeBlas:2019ehy,Biekoetter:2018ypq}. 
Such analyses are usually done by terminating the series in Eq.~\eqref{eq:intro:SMEFT-Lag} after dimension-6 operators. Yet, the need for precision calls for an investigation beyond ${\cal{O}}({1 / \LNP^2})$.
At the next non-trivial order, this includes studying the impact of dimension-8 SMEFT operators, but also double insertions of dimension-6 operators~\cite{Dawson:2021xei,Dawson:2022cmu,Corbett:2021eux,Boughezal:2022nof,Ellis:2022zdw,Alioli:2022fng,GomezAmbrosio:2022mpm,Heinrich:2022idm,Allwicher:2022gkm}.
An amplitude, $A_i$, for a lepton number conserving process can be parameterized in the SMEFT as a power series in ${1/ \Lambda^2}$,
\begin{align}
	\begin{split}
	A_i&\sim A_{i,\textrm{SM}} +\sum_j{C_j^6\over\Lambda^2}\alpha^6_{ij}+\sum_{j,k}{C_j^6C_k^6\over\Lambda^4}\alpha^{6^2}_{ijk} \\[-3pt]
	& +\sum_j{C_i^8\over\Lambda^4}\alpha^8_{ij} + \mathcal{O}{(1/\LNP^{6})} \, ,
	\end{split}
\end{align}
where the $\alpha$ coefficients are process dependent. The terms proportional to ${C_j^6C_k^6/\LNP^4}$
are the double insertions of interest here.  The amplitude-squared corresponding to a cross section is then expanded generically as,
\begin{align}
	|A_i|^2 &\sim  |A_{i,\textrm{SM}}|^2 
	+{1\over\LNP^2} \sum_j  2 \, \textrm{Re}{\left(A_{i,SM}^* C_j^6 \alpha^6_{ij}\right)} \nonumber \\ 
	 &\hspace{-5mm}+{1\over \Lambda^4} \bigg[
	\sum_{j,k} \bigg( C_j^6 C_k^{6*}\alpha^6_{ij}\alpha^{6*}_{ik} +2\,\textrm{Re}{\left(A_{i,\textrm{SM}}^* C_j^6 C_k^6\alpha_{ijk}^{6^2}\right)} \bigg) \nonumber \\
	&\eqspace + \sum_j 2 \, \textrm{Re}{\left( A_{i,\textrm{SM}}^* C_j^8\alpha_{ij}^8\right)} \bigg] + \mathcal{O}{(1/\LNP^{6})} \, .
	\label{eq:expand}
\end{align}
If a coefficient is well constrained by data, it may be sufficient to retain only the $\mathcal{O}(1/\Lambda^2)$ contributions to observables.  This is typically the case in fits to electroweak precision observables~\cite{daSilvaAlmeida:2018iqo,Dawson:2019clf,Berthier:2015oma}. However, for most of the SMEFT coefficients contributing to predictions for LHC observables, the $\mathcal{O}(1/\Lambda^4)$ terms play an important role.  Global fits~\cite{Ethier:2021bye,Biekoetter:2018ypq,Ellis:2020unq,DeBlas:2019ehy} include the first
term on the second line of Eq.~\eqref{eq:expand} (required to make the cross sections positive-definite), but the other terms of  ${\cal{O}}{({1/ \LNP^4})}$ are more subtle.  For tree-level processes, the second 
term on the second line of Eq.~\eqref{eq:expand} (which corresponds to a double insertion) is easily included~\cite{Brivio:2020onw,Degrande:2020evl}
and can have important numerical effects~\cite{Baglio:2020oqu}.  The dimension-8 contributions (first
term on the third line of Eq.~\eqref{eq:expand}) have been studied in only a few special cases and the numerical importance of these terms is not known in general~\cite{Hays:2018zze,Corbett:2021eux,Dawson:2022cmu,Dawson:2021xei}. In the case where the new physics that generates the SMEFT coefficients corresponds to a strongly interacting theory, it has been argued that the dimension-8 contributions are small~\cite{Contino:2016jqw}.

In the following, we present a preliminary investigation of the impact of double insertions on the inclusive gluon fusion Higgs boson production process.
This production channel has recently been calculated in the SM to $\textrm{N}^3\textrm{LO}$ QCD~\cite{Anastasiou:2016cez,Mistlberger:2018etf,Baglio:2022wzu}.
In the SMEFT, the NLO result with single insertions of dimension-6 operators is well known~\cite{Degrande:2012gr,Maltoni:2016yxb,Grazzini:2016paz,Deutschmann:2017qum,Harlander:2013oja}.
Gluon fusion Higgs production has also been calculated to all orders in $v^2/\LNP^2$ using the GeoSMEFT approach~\cite{Corbett:2021cil,Martin:2021cvs}.
Here, we present a study of the 1-loop contributions to the $g g \to h$ amplitude including all terms of $\order{1/(16 \pi^2\LNP^4)}$ and we investigate the numerical effects of double insertions of a consistent subset of dimension-6 SMEFT operators.

The paper is organized as follows. Section \ref{sec:smeft} contains a brief description of the SMEFT to $\order{1/\LNP^4}$.  
The 1-loop calculation of $gg\rightarrow h$ to 
$\order{1/(16\pi^2\LNP^4)}$ is presented in Section \ref{sec:calc}, including the insertion of two dimension-6 operators in the 1-loop amplitude
and the required counterterm for the $gg\rightarrow h$ process corresponding to the dimension-8
$(\varphi^\dagger
\varphi)^2G^{A,\mu\nu}G^B_{\mu\nu}$ operator.
Numerical effects of the double insertions are investigated in Section \ref{sec:results}, 
along with a discussion of
the potential effects of neglected contributions.
 Finally, we conclude in Section \ref{sec:conclusions}
 with a discussion of the path forward to a more complete study of the impact of $\mathcal{O}(1/\Lambda^4)$ effects.

\section{SMEFT to $\order{\LNP^{-4}}$}
\label{sec:smeft} 

We start by presenting the pieces of the 
dimension-6 SMEFT Lagrangian (in the Warsaw basis~\cite{Grzadkowski:2010es}) which are relevant for the calculation of the virtual 1-loop $gg\rightarrow h$ diagrams 
containing double insertions.
All the remaining necessary terms of the Lagrangian can be found in Ref.~\cite{Dedes:2017zog}. 
In the end of this section, we present the relationships up to ${\cal{O}}({1 / \LNP^4})$ between the original parameters of the Lagrangian and our input parameters~\cite{Hays:2018zze}.

We neglect finite contributions from dimension-8 terms. Although such contributions enter in the cross section at the same order as double insertions of dimension-6 operators, they can be treated separately, as they are not required to obtain a gauge-independent result. Yet, the  dimension-8 operators are in general required 
to absorb ultraviolet (UV) divergences of $\order{1/\LNP^4}$. 
There is a single dimension-8 operator that can be used to this end~\cite{Murphy:2020rsh,Li:2020gnx}, 
\begin{align}
	\label{eq:dim8-term}
\frac{C_{G^2 {\varphi}^4}}{\LNP^4} (\varphi^{\dagger} \varphi)^2 G_{\mu\nu}^{A} G^{A \mu\nu}.
\end{align}
When renormalizing the theory, the counterterm $\delta C_{G^2 \varphi^4}$ is generated from Eq.~\eqref{eq:dim8-term}. Below, we present the result
$\delta C_{G^2 \varphi^4}$ 
using minimal subtraction.
We work in minimal subtraction, which amounts to dropping all poles.  A complete understanding of dimension-8
renormalization in the SMEFT, including fermionic operators, does not yet exist, although significant progress has been made
in understanding the bosonic operators~\cite{DasBakshi:2022mwk,Chala:2021pll,Helset:2022pde,Chala:2021cgt,DeSousaFihaloGuedes:2022jti}. 

\subsection{Lagrangian and field redefinitions}

The relevant pieces of the dimension-6 SMEFT Lagrangian can be grouped into three terms,
\begin{align}
	\label{eq:SMEFT-Lag}
	\mathcal{L}_{\text{Higgs}} + \mathcal{L}_{\text{QCD}} + \mathcal{L}_{\text{fermions}} \, .
\end{align}
The first one is the Higgs Lagrangian,
\begin{align}
	\label{eq:Higgs-Lag}
	\begin{split}
		\mathcal{L}_{\text{Higgs}} &=
		\left(D^\mu \varphi\right)^{\dagger}
		 \left(D_\mu \varphi\right)+\mu^2 \varphi^{\dagger} \varphi 
		 - {\lambda\over 2}  \left(\varphi^{\dagger} \varphi\right)^2 \\
		&+ \frac{1}{\LNP^2} \Big[\ch\left(\varphi^{\dagger} \varphi\right)^3 
		+ \chbox \left(\varphi^{\dagger} \varphi\right) 
		\square\left(\varphi^{\dagger} \varphi\right) \\
		&\eqspace + \chd \left(\varphi^{\dagger} D^\mu \varphi\right)^* \left(\varphi^{\dagger} D_\mu \varphi\right)\Big] \, ,
	\end{split}
\end{align}
where $\varphi$ represents the Higgs doublet, which we parametrize as
\begin{align}
	\varphi=\left(\begin{array}{c}
		\varphi^{+} \\
		\frac{1}{\sqrt{2}}\left(v_T+h+i \varphi^0\right)
	\end{array}\right) \, .
\end{align}
Here, $v_T$ is the vacuum expectation value (vev) that minimizes the Higgs potential in the presence of the SMEFT operators, and $h$, $\varphi^{0}$ and $\varphi^{+}$ represent the Higgs, the neutral Goldstone, and the charged Goldstone boson fields, respectively.
The second term in Eq.~\eqref{eq:SMEFT-Lag} is the QCD Lagrangian, 
\begin{align}
\begin{split}
	\mathcal{L}_{\text{QCD}}& = -\frac{1}{4} G_{\mu\nu}^{A} G^{A \mu\nu} + \frac{\chg}{\LNP^2} \left(\varphi^{\dagger} \varphi \right) G_{\mu\nu}^{A} G^{A \mu\nu} \\
	&+\frac{C_G}{\LNP^2}f_{ABC}G_{\mu}^{A\nu} G_\nu^{B \rho}G_\rho^{C\mu}
	\, , \\
	\label{eq:LagQCD}
\end{split}
\end{align}
with
\begin{align}
	G_{\mu \nu}^A = \partial_\mu g_\nu^A - \partial_\nu g_\mu^A - g_s f^{A B C} g_\mu^B g_\nu^C,
\end{align}
where $g_{\mu }^A$ is the gluon field.
Finally, $\mathcal{L}_{\text{fermions}}$ is the fermionic Lagrangian,
\begin{align}
	\mathcal{L}_{\text{fermions}} &= -Y_u \, \bar{q}_L \varphi d_R  +\biggl[ \frac{\cht}{\LNP^2} \left( \varphi^{\dagger} \varphi \right) \left( \bar{q}_L  \tilde{\varphi} u_R  \right) + \mathrm{h.c.} \biggr]
	\nonumber \\
	&+\frac{\cll}{\LNP^2}({\overline {l}}_L\gamma_\mu l_L)({\overline {l}}_L^\prime\gamma_\mu l^\prime_L)\nonumber \\
	&+ \frac{\chlthree}{\LNP^2}\vpjt({\overline {l}}_L\tau^a \gamma_\mu l_L)
	\, ,
	\label{eq:Fermions-Lag}
\end{align}

with $q_L^{\text{T}} = \left(u_L, \ \ d_L \right)$, $l_L^{\text{T}} = \left(\nu_L, \ \ e_L \right)$,  $\tilde{\varphi} = i \sigma_2 \varphi^*$, and we retain only the top quark 
contributions.

To ensure that all fields have canonical kinetic terms, we need to perform the following shifts,
\begin{subequations}
\begin{align}
	h &\to h \, R_\phi^{-1} \, , \vphantom{g_{\mu}^A} \\
	\varphi^{0} &\to \varphi^{0} \, R_{\varphi^0}^{-1} \, ,  \\
	\label{eq:gluon-shift}
	g_{\mu }^A &\to g_{\mu}^A \, R_g^{-1} \, ,
\end{align}
\end{subequations}
where
\begin{subequations}
	\label{eq:field_shifts}
\begin{align}
	\label{eq:zhiggs}
	R_\varphi &= 1 - \frac{\vev^2}{\LNP^2} X_h - \frac{\vev^4}{2\LNP^4} X_h^2 + \order{\LNP^{-6}} \, , \\
	R_{\varphi^0} &= 1 + \frac{\vev^2}{4\LNP^2} \chd - \frac{\vev^4}{32\LNP^4} \chd^2 + \order{\LNP^{-6}} \, ,\\
	R_g &= 1 - \frac{\vev^2}{\LNP^2} \chg - \frac{\vev^4}{2\LNP^4} \chg^2 + \order{\LNP^{-6}} \, ,
\end{align}
\end{subequations}
with $X_h$ in Eq.~\eqref{eq:zhiggs} defined as
\begin{align}
	X_h \equiv \chbox - \frac{\chd}{4} \, .
\end{align}

\subsection{Input Parameters}

We choose as independent parameters
\begin{align}
	G_F \, , \ \alpha_s \, , \ M_Z \, , \ M_W \, , \ M_h \, , \ m_t \, ,
	\label{eq:indep_SM_params}
\end{align}
where $G_F$ is the Fermi constant, $\alpha_s$ is the strong coupling constant and $M_Z(M_W), M_h$ and $m_t$
are the gauge boson, Higgs and top masses.

The expression for $v_T$ can be determined through the amplitude for muon decay,
including double insertions of dimension-6 operators.
Assuming flavor universality of the WCs
\begin{align}
	G_F = \frac{1}{\sqrt{2} \, v_T^2} + \frac{\sqrt{2}}{\LNP^2} \left(\chlthree - \frac{1}{2} \cll\right)  + \frac{\vev^2}{\sqrt{2}} \frac{(\chlthree)^2}{\LNP^4} \, ,
\end{align}
which can be inverted to yield
\begin{align}
	\label{eq:vev_expression}
	\begin{split}
		v_T &= \frac{1}{ \sqtwoGf^\frac{1}{2} }
		+
		\frac{2 \chlthree - \cll }{2 \sqtwoGf^\frac{3}{2} \LNP^2} \\
		&+
		\frac{16 (\chlthree)^2 - 12 \, {\chlthree} \, {\cll} + 3 \cll^2}{8 \sqtwoGf^\frac{5}{2} \LNP^4}.
	\end{split}
\end{align}
The parameters $\mu^2$ and $\lambda$ are fixed by the 
requirement that the coefficient of the Higgs tadpole contribution 
vanishes (i.e. that $v_T$ is the true vev) and that the mass of the Higgs field in the Lagrangian is given by $M_h$. %
Using also Eq.~\eqref{eq:vev_expression}, we find
\begin{align}
	\begin{split}
	\label{eq:mu2}
	\mu^2 &= \frac{M_h^2}{2}
	+ \frac{3 \, {\ch} - 4 \sqrt{2}  X_h G_F \, M_h^2}{8 \, G_F^2 \, \LNP^2} \\
	& + \frac{ \big(2  \chlthree - \cll \big)  \big( 3  \sqrt{2} \ch -  
	4 X_h G_F  M_h^2 \big) }{8 \, G_F^3 \, \LNP^4} \, ,
	\end{split} \\[5pt]
	{\lambda} &= G_F  M_h^2\sqrt{2} + \frac{3 \sqrt{2}  {\ch} + 2 \big( \cll - 2 \chlthree -2 X_h  
		\big)  G_F M_h^2}{2 \, G_F \, {\Lambda}^2}  \nonumber \\
	&- \frac{3 \, {\ch} \,  \big( \cll - 2 \, \chlthree\big)  + \sqrt{2} \, 
	(\chlthree)^2 \, G_F \, M_h^2}{2 \, G_F^2 \, {\Lambda}^4} \, .
\label{eq:lambda}
\end{align}

\begin{figure*}[t]
	\centering
	\includegraphics[width=105pt]{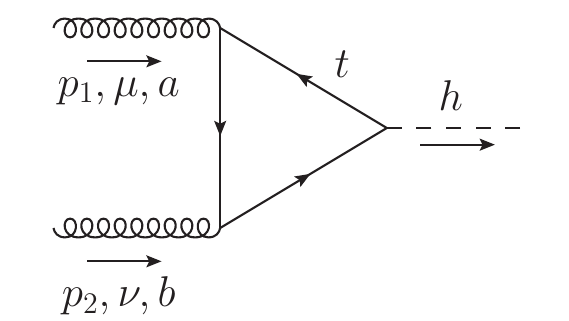} \hspace{-12pt}
	\includegraphics[width=105pt]{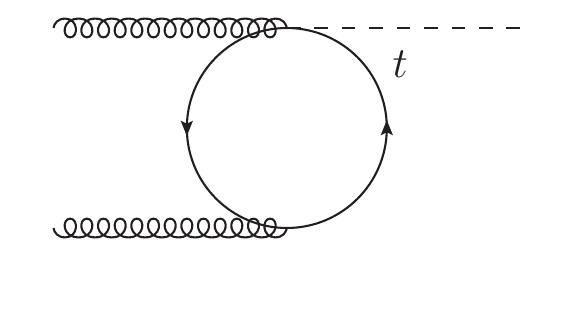} \hspace{-12pt}
	\includegraphics[width=105pt]{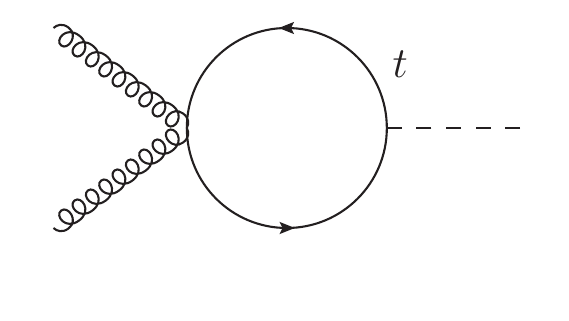} \hspace{-12pt}
	\includegraphics[width=105pt]{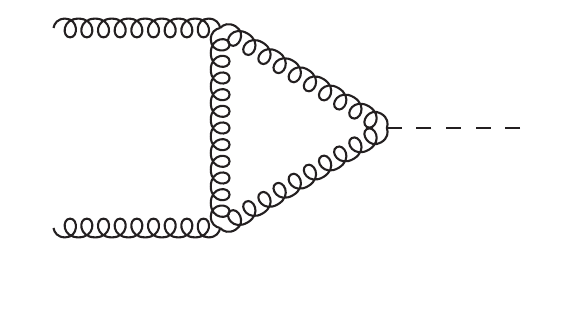} \hspace{-12pt}
	\includegraphics[width=105pt]{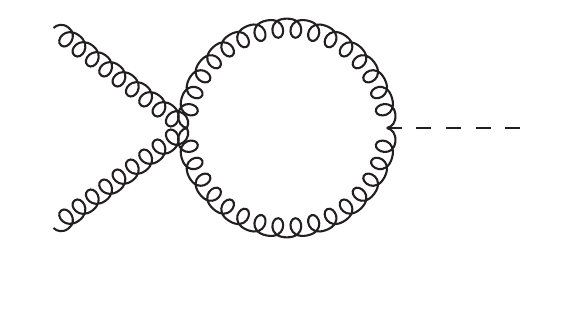} \\[-10pt]
	$(1)$ \hspace{80pt} $(2)$ \hspace{80pt} $(3)$ \hspace{80pt} $(4)$ \hspace{80pt} $(5)$ \\[10pt]
	\includegraphics[width=105pt]{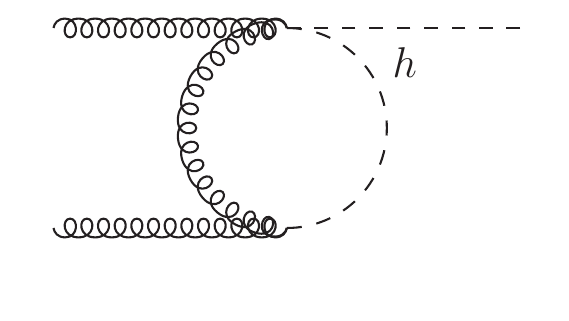} \hspace{-12pt}
	\includegraphics[width=105pt]{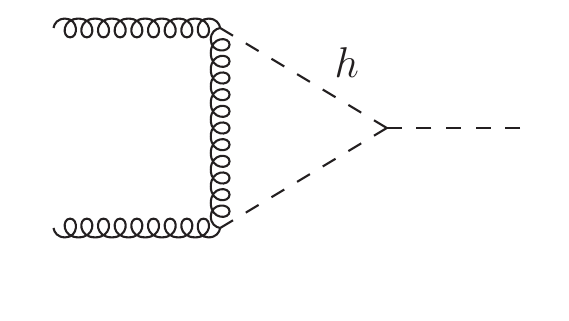} \hspace{-12pt}
	\includegraphics[width=105pt]{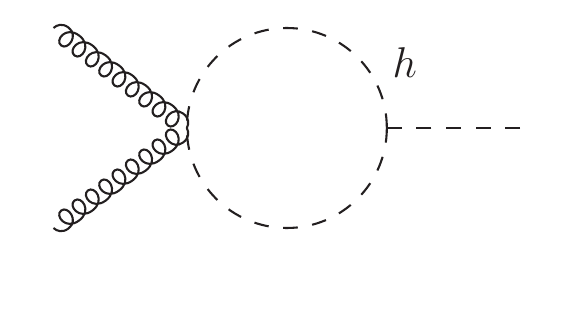} \hspace{-12pt}
	\includegraphics[width=105pt]{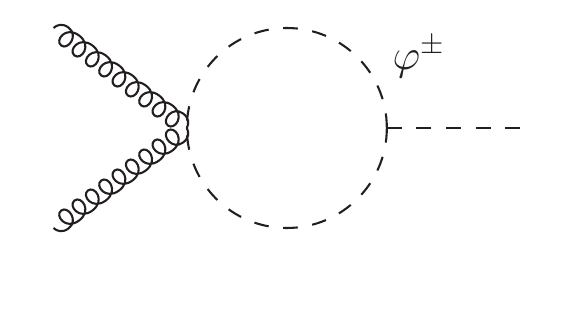} \hspace{-12pt}
	\includegraphics[width=105pt]{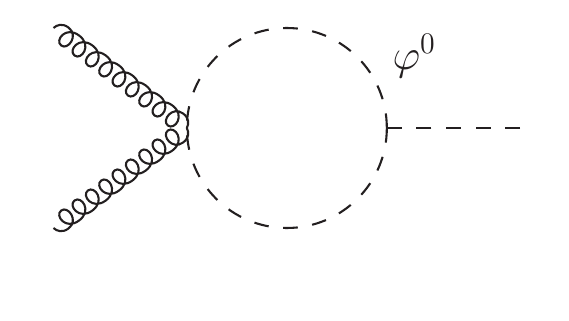} \\[-10pt]
	$(6)$ \hspace{80pt} $(7)$ \hspace{80pt} $(8)$ \hspace{80pt} $(9)$ \hspace{80pt} $(10)$ \\[10pt]
	\caption{Virtual 1-loop contributions to the gluon fusion to Higgs amplitude including contributions from 
		both single and double insertions of dimension-6 SMEFT operators. Conventions used throughout the paper concerning 4-momenta, Lorentz indices and colour indices are shown in diagram $(1)$. Note that diagrams $(1)$, $(2)$ and $(6)$ also contribute with crossed initial states (not shown for compactness).}
	\label{fig:ampl:diags}
\end{figure*}
The top quark Yukawa coupling is determined by requiring that the mass of the top-quark field in Eq.~\eqref{eq:Fermions-Lag} is given by $m_t$, 
\begin{align}
	Y_t &= \sqrt{2} \, \sqtwoGf^{\frac{1}{2}} \, m_t  \nonumber \\
	&\eqspace \times \bigg[1 - \frac{2 {\chlthree} - {\cll}}{2 \sqtwoGf \LNP^2} - \frac{8 (\chlthree)^2 - 4  \chlthree 
		{\cll}+\cll^2}{8 \sqtwoGf^2 \LNP^4} \bigg] \nonumber \\
	&+\frac{\cht}{2 \sqtwoGf \LNP^2} \bigg[ 1 +   \frac{2 {\chlthree} - {\cll}}{ \sqtwoGf \LNP^2} \bigg] \, .
\end{align}
Finally, $g_s^2$ can be related to $4 \pi \alpha_s$ through the inverse transformation of Eq.~\eqref{eq:gluon-shift}
and we find
\begin{align}
\begin{split}
g_s &=
\bar{g}_s \Bigg[1
- \dfrac{1}{\sqrt{2}G_F} \frac{\chg}{\LNP^2} \\
&\eqspace - \frac{1}{4 G_F^2} \frac{\chg \, (\chg + 4 \chlthree - 2 \cll)}{\LNP^4} \Bigg] \, ,
\end{split}
\end{align}
where we defined
\begin{equation}
\bar{g}_s \equiv \sqrt{4 \pi \alpha_s} \, .
\end{equation}

\begin{table*}[t]
	\begin{center}
		\begin{tabular}{lrclccrclccr}
			\cmidrule[1pt]{1-2} \cmidrule[1pt]{4-7} \cmidrule[1pt]{9-12}
			${10^{-4}\over \text{GeV}^2}  \cdot a_i$ \hspace{-10pt} & linear & & ${10^{-10}\over \text{GeV}^4} \cdot b_{ij}$ \hspace{-8pt} & single & double & ratio & & ${10^{-10}\over \text{GeV}^4} \cdot b_{ij}$ \hspace{-8pt} & single & double & ratio  \\
			\cmidrule{1-2} \cmidrule{4-7} \cmidrule{9-12}
			$\chlthree$ & -12.13 & & $\chlthree^2$      &   0.3678 & -0.3678 &  -1 & & $\cht \chlthree$   &   0.7447 & -0.7447 &  -1 \\
			\cmidrule{4-7}
			$\cll$ & 6.06 & & $\cll \chlthree$   &  -0.3678 &         - &   - & &  $\cht \cll$        &  -0.3723 &  0.3723 &  -1 \\[5pt]
			$\chbox$ & 12.13 & & $\cll^2$           &  0.0919 &         - &   - & &  $\cht \chbox$      &  -0.7447 &    -1.4893 & 1/2 \\
			\cmidrule{4-7}
			$\chd$ & -3.03 & & $\chbox \chlthree$ &  -0.7355 &         - &   - & &  $\cht \chd$        &   0.1862 &  0.3723 & 1/2  \\[5pt]
			$\cht$ & -12.28 & & $\chbox \cll$      &   0.3678 &         - &   - & & $\cht^2$           &    0.3769 &   0.3769 &   1  \\
			\cmidrule{9-12}
			$\cug$ & 19.35 & & $\chbox^2$         &   0.3678 &   1.4711 & 1/4 & & $\cug \chlthree$ & -1.1732 & -1.1732 & 1  \\
			\cmidrule[1pt]{1-2} \cmidrule{4-7} 
			&  & & $\chd \chlthree$   &   0.1839 &         - &   - & &   $\cug \cll$      & 0.5866 & 0.5866 & 1 \\[5pt]
			&  & & $\chd \cll$        & -0.0919 &         - &   - & &   $\cug \chbox$    & 1.1732 & 2.3465 & 1/2  \\[5pt]
			&&&$\chd \chbox$      &  -0.1839 & -0.7355 & 1/4 & & $\cug \chd$      & -0.2933 & -0.5866 & 1/2 \\[5pt]
			&&&$\chd^2$           &  0.0230 & 0.0919 & 1/4 & & $\cug \cht$      & -1.1878 & -0.0661 & 17.97  \\
			\cmidrule{4-7} 
			&&& & & & && $\cug^2$         & 0.9357 & 1.3909 & 0.6727  \\
			\cmidrule[1pt]{9-12}
		\end{tabular}
		\caption{Numerical results for linear coefficients $a_i$ and coefficients $b_{ij}$ of pairs of SMEFT WCs, c.f. Eq.~\eqref{eq:res:amplsq}. Results are shown with (third column) or without (second column) double insertions. In the fourth column we show the ratio of \textit{single} coefficients over \textit{double} coefficients. Ratios given as rational numbers are exact. Numerical values for physical parameters are reported in section~\ref{sec:results}. See text for further details.}
		\label{tab:results:numeric}
	\end{center}
\end{table*}

\section{Calculation}
\label{sec:calc}

We now describe the 1-loop calculation of the $gg\rightarrow h$ amplitude to 
$\order{1/(16\pi^2\LNP^4)}$.
The Feynman rules accurate to ${\cal{O}}({1/\Lambda^4})$ 
that are relevant for our calculation are given in  Appendix \ref{app:feynman}.
Lorentz and gauge invariance imply that at any order, the amplitude for
$g^A(p_1^{\mu}) g^B(p_2^{\nu})\rightarrow h$ must have the form, 
\begin{align}
\label{eq:fdef}
    A^{\mu\nu}(p_1,p_2)=i\delta_{AB}\biggl( {p_1^\nu p_2^\mu-p_1\cdot p_2}g^{\mu\nu}\biggr) \sum_i F_i \, , 
\end{align}
where, up to 1-loop, 
\begin{align}
\label{eq:Fs}
   \sum_i F_i = F_0 + F_{\text{V}} + F_{\text{CT}} \, ,
\end{align}
with
$F_0$ representing the tree-level SMEFT contribution, 
$F_{\text{V}}$ the virtual 1-loop amplitude and
$F_{\text{CT}}$ the  total counterterm.

The tree-level contribution  is given by
\begin{align}
    \label{eq:tree-rule-var}
    \begin{split}
    F_0 &=
    \dfrac{4{\chg}}{(\sqrt{2}G_F)^{\frac{1}{2}} \Lambda^2}
    + \dfrac{\chg}{(\sqrt{2}G_F)^{\frac{3}{2}}  \Lambda^4} \\
    &\eqspace \times \bigg[ 8 \, \chg + 4 X_h +
      4 \chlthree - 2 \cll \bigg]
    \, .
    \end{split}
\end{align}
$F_{\text{V}}$ is computed from the diagrams shown in Fig. \ref{fig:ampl:diags}, using the software \ts{FeynMaster}~\cite{Fontes:2019wqh,*Fontes:2021iue,Christensen:2008py,*Alloul:2013bka,Nogueira:1991ex,Mertig:1990an,*Shtabovenko:2016sxi,*Shtabovenko:2020gxv,*Shtabovenko:2016whf}.
We use the true vev up to 1-loop order~\cite{Fontes:2021kue} and we work in the Parameter Renormalized tadpole scheme~\cite{Denner:1991kt,*Denner:2018opp}.
Analytic results for $F_\textrm{V}$ can be found in the auxiliary file submitted with this paper. 
Finally, $F_{\text{CT}}$ is determined by identifying the original parameters and fields in Eqs~(\ref{eq:dim8-term},~\ref{eq:SMEFT-Lag})  as bare parameters (with index ``$\scriptstyle (0)$'') and 
by expanding them into  renormalized quantities,
\begingroup
\allowdisplaybreaks
\begin{subequations}
\label{eq:renorm-expansions}
\begin{align}
    h_{\scriptscriptstyle (0)} &= \Big(1 + \dfrac{1}{2} \delta Z_h \Big) h \, , \\
    g_{\scriptscriptstyle(0)}^{A, \mu} &= \Big(1 + \dfrac{1}{2} \delta Z_{g} \Big) g^{A,\mu} \, , \\
    \label{eq:GF-ren}
    G_{F\scriptscriptstyle(0)} &= \left(1 + \delta G_F \right) G_F \, , \vphantom{\frac{1}{2}} \\
    C_{X \scriptscriptstyle (0)} &= C_X + \delta C_X \, , \vphantom{\frac{1}{2}}
\end{align}
\end{subequations}
\endgroup
where $C_X$ represents a generic WC. The expression for $F_{\text{CT}}$ is given in Appendix \ref{app:counter}.\footnote{As discussed in Section \ref{sec:smeft}, we ignore finite effects from dimension-8 operators (i.e. we set the renormalized WC
$C_{G^2\varphi^4}$ to zero).}

This allows us to determine $\delta C_{G^2\varphi^4}$ by requiring Eq. \eqref{eq:Fs} be free from divergences.
We work in dimensional regularization, using $D = 4 - 2\epsilon$ for the spacetime dimension, and fix the counterterms of the WCs in the minimal subtraction scheme~\cite{tHooft:1973mfk,*Weinberg:1973xwm}.  We perform the calculation in two independent ways: {\textit{i)} we subtract known infrared (IR)  poles using results of Ref.~\cite{Catani:1998bh}; and \textit{ii)}
we  use \ts{Package-X}~\cite{Patel:2015tea,*Patel:2016fam} and consider only UV poles.

It is sufficient to compute the counterterms in Eq.~\eqref{eq:rule-for-CT}  to order ${\cal{O}}(1/\Lambda^2)$, since 
Eq.~\eqref{eq:rule-for-CT}  is already $ {\cal{O}}(1/\Lambda^2)$.
 $\delta Z_h$ and $\delta Z_g$ can be computed from the Higgs and gluon self energies at 1-loop, respectively; explicit expressions can be found in Appendix~\ref{app:counter}.
$\delta G_F$ is given by
\begin{align}
	\label{eq:deltaGF}
	\begin{split}
		\delta G_F &= 
		- \dfrac{1}{16 \pi^2} \dfrac{G_F}{\sqrt{2}} {\Delta r}_{\text{SM}} \\
		&- \dfrac{1}{16 \pi^2} \dfrac{1}{\Lambda^2} {\Delta r}_{\mathrm{EFT}} + \dfrac{1}{2}\left( 2\chlthree -  \cll \right) \dfrac{{\Delta r}_{\text{SM}}}{16 \pi^2 \Lambda^2} \\
		&+  \dfrac{1}{2} \left( 2\delta \chlthree - \delta \cll\right) \dfrac{\sqrt{2}}{G_F \Lambda^2},
	\end{split}
\end{align}
where the expressions for ${\Delta r}_{\text{SM}}$ and ${\Delta r}_{\text{EFT}}$ can be found in  Appendix D of Ref.~\cite{Dawson:2018pyl}.
The contributions from $\delta \chlthree$ and $\delta \cll$ cancel when Eq.~\eqref{eq:deltaGF} is used in Eq.~\eqref{eq:rule-for-CT}. 
The contribution to $\delta\chg$ of ${\cal{O}}(1/\Lambda^2)$
 can be obtained from Refs~\cite{Jenkins:2013zja,*Jenkins:2013wua,*Alonso:2013hga}; we confirmed their result by
requiring that Eq. \ref{eq:fdef} be finite to ${\cal{O}}(1/\Lambda^2)$ and present it in Eq.~\eqref{eq:deltachg}. Combining these elements, we find the expression for $\delta C_{G^2 \varphi^4}$ given in Eq.~\eqref{eq:ct8_final}.

\section{Impact of double insertions}
\label{sec:results} 

To study the impact of double insertions on the 1-loop amplitude of the gluon fusion process, we compute the amplitude squared in two ways: \textit{i)} we truncate the amplitude at   ${\cal{O}}(1/\LNP^2)$ and then compute the amplitude squared; \textit{ii)} we compute the amplitude to ${\cal{O}}(1/\LNP^4)$  and then truncate the amplitude squared at  ${\cal{O}}(1/\LNP^4)$.  
The first truncation is not sensitive to the double insertions of the dimension-6 operators, and we label it as 
\textit{``single''}.
The second truncation is sensitive to the double insertions of SMEFT operators, and we label it as \textit{``double''}.
We note that the latter is in fact a complete computation of the virtual amplitude up to $\order{1/\LNP^4}$ at 1-loop,
neglecting finite contributions from dimension-8 operators.
Since the WC $\chg$ contributes at tree-level, the double insertions proportional to $\chg$ require the computation of 2-loop virtual graphs with single insertions of dimension-6 operators, along
with 1-loop virtual graphs proportional to $\chg$ to obtain an IR finite result.

\begin{figure}[t]
	\centering
	\hspace{-20pt}\includegraphics[width=200pt]{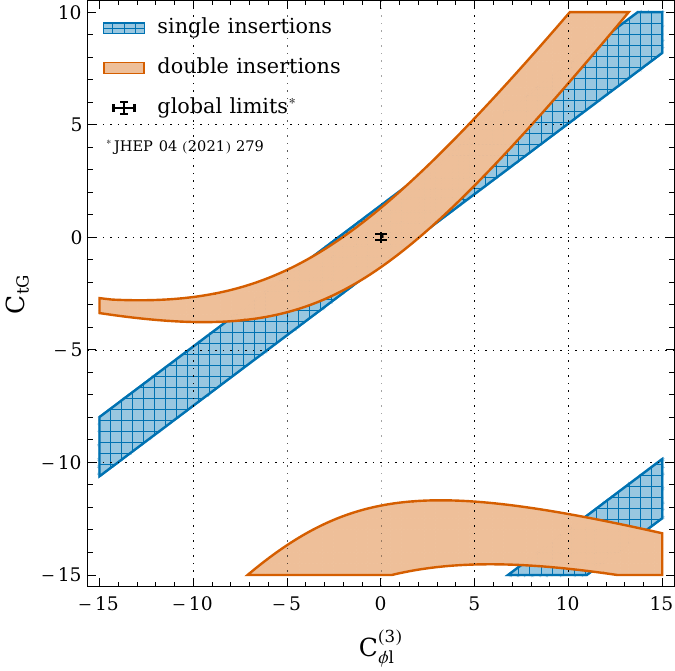}  \\[10pt]
	\hspace{-15pt}\includegraphics[width=196pt]{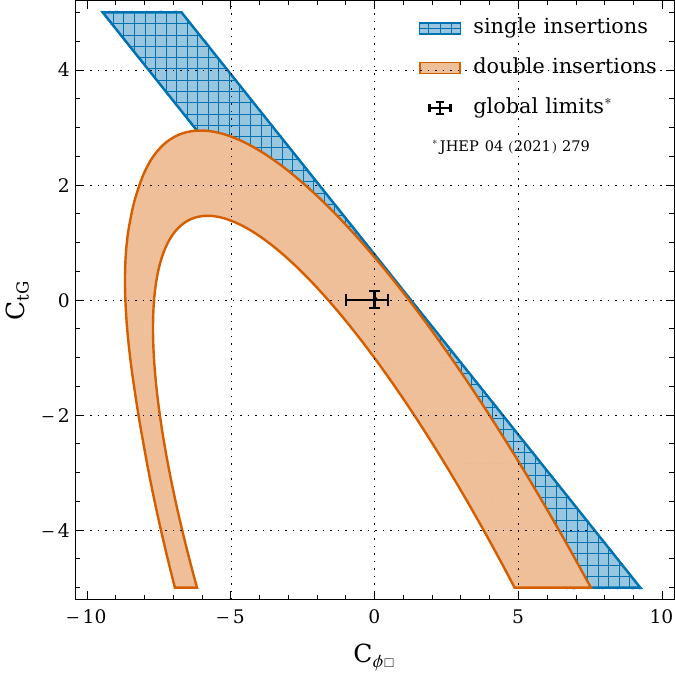} 
	\caption{ Regions where $| \mu_{ggh}-1|  < 5\%$ are shown for single insertions (squared blue) and double insertions (orange). The limits from global fits to individual operators at $95\%$ CL are denoted by the black cross.~\cite{Ellis:2020unq,Ethier:2021bye,Biekoetter:2018ypq}. The  WCs not shown are varied over values allowed by the 95$\%$ CL fits to individual coefficients of Ref.~\cite{Ethier:2021bye}.}
	\label{fig:chbox/chl3_ctg}
\end{figure}

As a first step in understanding the relevance of double
insertions, we consider a scenario where $\chg$ is generated at loop level and thus can be consistently set to zero
after renormalization.  This is a realistic scenario from a model building point of view.  At tree-level, scalars,
vector-like quarks,
and vector particles in arbitrary representations that contribution to the dimension-6 SMEFT Lagragian  do not generate  $\chg$ contributions~\cite{deBlas:2017xtg}.
It is interesting to note that vector-like quarks generate $\chg$ at 1-loop consistent with our assumption.  When we set  $\chg=0$, there are no real corrections and we can study the numerical effects of the double insertions 
from the remaining operators using our finite results for the renormalized amplitude
to construct a cross section normalized to the SM result.\footnote{We have explicitly checked the gauge independence of our results.}

For the numerical results reported below, we use   $M_h = 125$ GeV, $M_W = 80.377$ GeV ,
$M_Z = 91.1876$ GeV, $m_t = 172$ GeV, $G_F = 1.166 \cdot 10^{-5}$ GeV${}^{-2}$ and  $\alpha_s = 0.1179$.  The renormalization scale $\mu$ is chosen to be equal to the Higgs mass $M_h$.
Finally, we write the virtual amplitude squared as,
\begin{align}
\label{eq:res:amplsq}
\left| {\sum_i F_i \over F_\textrm{SM}}\right|^2\equiv 1+\sum_i a_i \, {C_i\over\Lambda^2}+\sum_{i, \, j \le i} b_{ij} \, {C_i C_j\over\Lambda^4} \, .
\end{align}
In the $\chg=0$ limit that we are working in,
\begin{align}
\mu_{ggh}&\equiv {\sigma(gg\rightarrow h)\over \sigma(gg\rightarrow h)|_{\text{SM}} }=
\left| {\sum_i F_i \over F_\textrm{SM}}\right|^2\, .
\end{align}
Numerical results for $a_i$ and $b_{ij}$ in the 2 expansions
 at ${\cal{O}}(1/\LNP^4)$ are presented in Table~\ref{tab:results:numeric}.

We first note that some contributions that contain $\cll$ or $\chlthree$ are present in the \textit{single}  but vanish in the \textit{double} setup.
From the Feynman diagrams shown in Fig.~\ref{fig:ampl:diags} it can be easily seen that these contributions are proportional to $1/(R_\varphi^2v_T^2)$, which vanishes in the \textit{double} expansion.
Consequently, the functional dependence of the amplitude on these WCs in the two expansions is quite different; for example, we show this for the combination of $\chlthree$ and $\cug$ in the upper plot in Fig.~\ref{fig:chbox/chl3_ctg}. 
In this figure we show the regions where $| \mu_{ggh}-1|$  is less than $5\,\%$.
For a given
value of $\chlthree$ and $\cug$, the remaining coefficients $\cll$, $\chbox$, $\chd$, and $\cht$ are varied over the region allowed by the $95\%$ CL individual fits of Ref.~\cite{Ellis:2020unq}.\footnote{Limits used in all figures for WCs not shown explicitly are 
	$-0.5 \le 10^2 \cdot\cll \le 1.9$, 
	$ -1.0 \le 10^2\cdot\chlthree \le 0.3 $, 
	$-1.0 \le \chbox \le 0.5$, 
	$-2.3 \le 10^2 \cdot \chd \le 0.3$, 
	$-1.0 \le \cht \le 0.8$ and 
	$-1.3 \le 10 \cdot \cug \le 1.5$.}
It is clear that the difference between the \textit{single} and \textit{double} insertion expansions has
no phenomenological relevance, since the values of the parameters plotted are excluded by fits to Higgs data~\cite{Ellis:2020unq,Ethier:2021bye,Biekoetter:2018ypq}. 
We do not show it explicitly, but we have checked that the same conclusion holds for all other combinations that include $\cll$ and/or $\chlthree$.

We also observe a non-trivial change in the coefficient of $\cug$ and we show a fit in combination with $\chbox$ to the value of the SM amplitude squared in Fig.~\ref{fig:chbox/chl3_ctg}~(bottom).
Also in this case, significant differences between \textit{single} and \textit{double} expansions only occur for values of the WCs far beyond current single parameter limits~\cite{Ellis:2020unq}.

The biggest change is in the coefficient of $\cug\cht$.
For this combination of WCs, the allowed parameter space is available in Ref.~\cite{Ellis:2020unq} from 
 2-parameter fits to Higgs and Higgs plus  top data at $95\,\%$ CL.
\begin{figure}[]
	\centering
	\hspace{-20pt}\includegraphics[width=196pt]{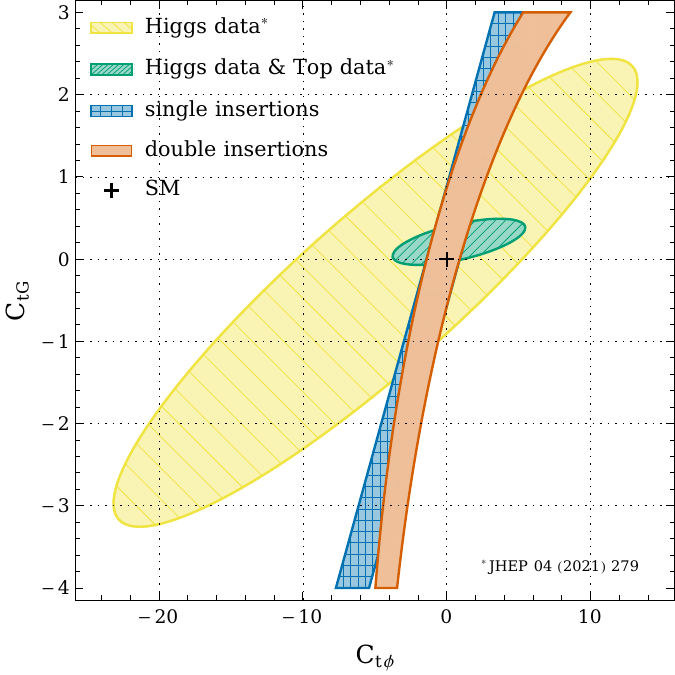}
	\caption{Allowed parameter space from a 2-parameter fit to $\cht$ and $\cug$. Yellow (hashed) and green (fine hashed) ellipses show constraints from linear fits at $95\%$ CL to Higgs data and Higgs plus  top data respectively~\cite{Ellis:2020unq}. Regions where $| \mu_{ggh}-1| < 5\%$  are shown for single insertions (squared blue) and double insertions (orange). The WCs not shown are varied over values allowed by the 95$\%$ CL fits to individual coefficients of Ref.~\cite{Ethier:2021bye}.
	}
	\label{fig:ctg_cth}
\end{figure}
In Fig.~\ref{fig:ctg_cth}, we show these regions together with a fit to $| \mu_{ggh}-1 | < 5\%$.
The difference in the results for \textit{single} and \textit{double} expansions is small and
demonstrates the power of including top data in the fits.
While  fits to  Higgs data alone show a small sensitivity to the expansion,
when top data is included with the Higgs data, there is again no difference between the two expansions in the region
allowed by global fits.\footnote{We stress that we have not included unknown dimension-8 contributions that could also contribute at $\mathcal{O}(1/\LNP^4)$.}


\section{Conclusions}
\label{sec:conclusions}

We computed the 1-loop amplitude for the gluon fusion process $gg\rightarrow h$ including all contributions of dimension-6 operators up to ${\cal{O}}({1/(16 \pi^2 \Lambda^4)})$.
This includes double insertions of dimension-6 operators and the  relationships between parameters in the SMEFT Lagrangian and physical observables to this order.
We derived the necessary Feynman rules that are valid up to ${\cal{O}}(1/\LNP^4)$ and determined the required counterterm to obtain a UV finite result at this order.
For our numerical studies, we considered  the limit $\chg = 0$ which ensures that there are no infrared singularities.
We note that this is a well motivated scenario, since in many BSM models $\chg$ is only generated at 1-loop level.
We then compared the gluon fusion cross section
in different expansions up to  $\order{1/\LNP^4}$ and found that the impact of the double insertions is 
negligible for values of the WCs allowed by global fits and neglecting the unknown dimension-8 contributions. 

An extension of this study including the effects of $\chg$ and double insertions would require
2-loop virtual amplitudes with up to two insertions of dimension-6 SMEFT operators as well as real-virtual and double real emission contributions.
We leave this exercise for future  investigations.  

Digital data associated with this research is contained in the auxiliary file attached to this paper.

\begin{acknowledgments}
	We thank Pier Paolo Giardino, Guilherme Guedes, Matt Sullivan and Robert Szafron for useful discussions. The research of KA, SD and DF is supported by the United States Department of Energy under Grant Contract DE-SC0012704.
\end{acknowledgments}


\appendix
\onecolumngrid

\section{Counterterms}
\label{app:counter}

Here, we collect results related to the renormalization. In what follows, all results are written in the Feynman gauge and, unless explicitly stated otherwise, $\epsilon$ represents $\epsilon_{\text{UV}}$ (i.e. a UV pole).

The counterterm $\delta C_{\varphi G}$ receives contributions of ${\cal{O}}(1/\Lambda^4)$.  The bosonic
contributions of ${\cal{O}}(1/\Lambda^4)$ are given in Ref.~\cite{Helset:2022pde} , while while the fermionic contributions are unknown.  We denote the total 
${\cal{O}}(1/\Lambda^4)$ contribution to $\delta C_{\varphi G}$ as $\delta C_{\varphi G}^8$
and the ${\cal{O}}(1/\Lambda^2)$ contribution as  $\delta C_{\varphi G}^6$,
\begin{eqnarray}
	\label{eq:deltachg}
		\epsilon \, \, \delta C_{\varphi G} 
		&=&\epsilon\biggl[ \delta C_{\varphi G}^6+{1 \over \sqrt{2} G_F}{\delta C_{\varphi G}^8\over \LNP^2}\biggr]
		\nonumber \\
		&=& -  \dfrac{\sqrt{\alpha_s \, G_F} \, m_t}{2^{\frac{5}{4}} \pi^{\frac{3}{2}}} \, \cug  + 
		\dfrac{3 \sqrt{2} G_F (M_h^2 + 2 m_t^2 - 2 M_W^2 - M_Z^2) - 28 \pi \alpha_s}{16 \pi^2} \, \chg 
		+\epsilon \, {1 \over \sqrt{2} G_F}{\delta C_{\varphi G}^8\over\LNP^2}
		\, .
\end{eqnarray}
The quantity $F_{\text{CT}}$ defined in Eq. \ref{eq:Fs} is given by
\begingroup
\allowdisplaybreaks
\begin{align}
    F_{\text{CT}}
	&= 
	\dfrac{2}{(\sqrt{2} G_F)^{\frac{1}{2}}}
	\dfrac{2 \, {\delta C_{\varphi G}^6} + {C_{\varphi G}} \,  \left(\delta Z_h + 2 \, {\delta Z_g} - {\delta G_F} \right)}{{\Lambda}^2} 
	+\dfrac{4}
	{(\sqrt{2} G_F)^{\frac{3}{2}}}
	\dfrac{{\delta C_{\varphi G}^8}}{\LNP^4}
	\nonumber \\
	& + \dfrac{1}{2 (\sqrt{2} G_F)^{\frac{3}{2}}  \, {\Lambda}^4} \bigg[ 8 \, {\delta C_{G^2 \varphi^4}} + 8 \, {C_{\varphi \square}} \, {\delta C_{\varphi G}^6} - 2 \, {C_{\varphi D}} \, {\delta C_{\varphi G}^6} + 32 \, {C_{\varphi G}} \, {\delta C_{\varphi G}^6} + 8 \, {C_{\varphi l}^{(3)}} \, {\delta C_{\varphi G}^6} - 4 \, {C_{ll}} \, {\delta C_{\varphi G}^6}  + 8 \, {C_{\varphi G}} \, {\delta C_{\varphi l}^{(3)}} \nonumber \\
	&  - 4 \, {C_{\varphi G}} \, {\delta C_{ll}} + 3 \, {C_{\varphi D}} \, {C_{\varphi G}} \, {\delta G_F} - 24 \, C_{\varphi G}^2 \, {\delta G_F}  - 12 \, {C_{\varphi G}} \, {C_{\varphi l}^{(3)}} \, {\delta G_F} + 6 \, {C_{\varphi G}} \, {C_{ll}} \, {\delta G_F} - 2 \, {C_{\varphi D}} \, {C_{\varphi G}} \, {\delta Z_g} \nonumber \\
	& + 16 \, C_{\varphi G}^2 \, {\delta Z_g} + 8 \, {C_{\varphi G}} \, {C_{\varphi l}^{(3)}} \, {\delta Z_g}  - 4 \, {C_{\varphi G}} \, {C_{ll}} \, {\delta Z_g} - {C_{\varphi D}} \, {C_{\varphi G}} \, {\delta Z_h} + 8 \, C_{\varphi G}^2 \, {\delta Z_h}  + 4 \, {C_{\varphi G}} \, {C_{\varphi l}^{(3)}} \, {\delta Z_h} \nonumber \\
	&  - 2 \, {C_{\varphi G}} \, {C_{ll}} \, {\delta Z_h} 
	+ 4 \, {C_{\varphi \square}} \, {C_{\varphi G}} \,  \left( -3 \, {\delta G_F} + 2 \, {\delta Z_g} + {\delta Z_h} \right) \bigg].
	\label{eq:rule-for-CT}
\end{align}
\endgroup
The poles of $\delta Z_h$ and $\delta Z_g$ are respectively such that
\begingroup
\allowdisplaybreaks
\begin{align}
\label{eq:deltaZhUV}
\begin{split}
\epsilon \,\, \delta {Z_h}\big|_{\text{poles}}
&=
\dfrac{2 \, M_W^2 - 3 \, m_t^2+ M_Z^2}{4 \sqrt{2} \pi^2} G_F + \dfrac{1}{\Lambda^2}
\bigg[
\dfrac{3 \, (M_Z^2-M_W^2)}{4 \pi^2} \, C_{\varphi B} + 
\dfrac{4 \, M_W^2 + 2 \, M_Z^2 - 7 \, M_h^2 - 6 \, m_t^2}{8 \pi^2} \, C_{\varphi \square}  \\
& + \dfrac{5 \, M_h^2 + 6 \, m_t^2 - 4 \, M_W^2 + M_Z^2}{32 \pi^2} \, C_{\varphi D}
+ \dfrac{3 \, m_t^2 - 2 \, M_W^2 - M_Z^2}{4 \pi^2} \, C_{\varphi l}^{(3)} + \dfrac{9 \, M_W^2}{4 \pi^2} C_{\varphi W}   \\
& + \dfrac{3 M_W \sqrt{M_Z^2-M_W^2}}{4 \pi^2} C_{\varphi WB}
+ \dfrac{M_Z^2 + 2 \, M_W^2 - 3 \, m_t^2}{8 \pi^2} C_{ll}
+ \dfrac{3 m_t}{4 \sqrt{2} \pi^2
(\sqrt{2} G_F)^{\frac{1}{2}}} C_{t \varphi}
\bigg] \, ,
\end{split} \\[10pt]
\delta Z_g\big|_{\text{poles}}
&= 
- \dfrac{5 \, \alpha_s}{12 \, \pi \, \epsilon_{\text{IR}}}
+
\dfrac{1}{\epsilon_{\text{UV}}}
\Bigg\{
\dfrac{\alpha_s}{4 \pi}
+ 
\dfrac{1}{\Lambda^2}
\bigg[
\frac{\sqrt{\alpha_s} \, m_t}{\sqrt{2} \pi^{\frac{3}{2}} (\sqrt{2} G_F)^{\frac{1}{2}}} C_{tG} 
- \frac{M_h^2 + 2 M_W^2 + M_Z^2}{8 \pi^2} C_{\varphi G} 
\bigg]
\Bigg\} \, .
\end{align}
\endgroup
Finally, the counterterm $\delta C_{G^2 \varphi^4}$ is
\begingroup
\allowdisplaybreaks
\begin{align}
\label{eq:ct8_final}
	\epsilon \, \, \delta C_{G^2 \varphi^4}
	&= \, \, \,
	C_{\varphi G}^2 \, \Bigg\{ \dfrac{3 \, \sqrt{2} \, {G_F} \, M_h^2 + 28 \, \alpha_s \, \pi}{8 \,  \, \pi^2} \Bigg\}
	+ C_{\varphi G} \, 
	\Bigg\{
	-\dfrac{3}{16 \, \pi^2} \, {C_{H}}
	- \dfrac{3 \, (\sqrt{2} G_F)^{\frac{1}{2}} {m_t}}{8 \, \sqrt{2} \, \pi^2} \, {C_{tH}}
	- \dfrac{9 \, {G_F} \, M_W^2}{4 \, \sqrt{2} \, \pi^2} \, C_{\varphi W} \nonumber   \\
	& + \dfrac{3\, {G_F} \,  \left( m_t^2 - 2 \, M_W^2 \right) }{4 \, \sqrt{2} \, \pi^2}  \, C_{\varphi q}^{(3)} 
	+ \dfrac{3\, {G_F} \,  \left( {M_W} - {M_Z} \right)  \,  \left( {M_W} + {M_Z} \right) }{4 \, \sqrt{2} \, \pi^2}  \, {C_{\varphi B}} - \dfrac{3\, {G_F} \, {M_W} \, \sqrt{M_Z^2 -M_W^2 }}{4 \, \sqrt{2} \, \pi^2} \, {C_{\varphi WB}} \nonumber  \\
	& + \dfrac{{G_F} \,  \left( 45 \, M_h^2 + 36 \, m_t^2 - 46 \, M_W^2 - 18 \, M_Z^2 \right) }{24 \, \sqrt{2} \, \pi^2} \, {C_{\varphi \square}} 
	+ \dfrac{3\, {G_F} \,  \left( M_h^2 + 2 \, m_t^2 - 2 \, M_W^2 - M_Z^2 \right) }{8 \, \sqrt{2} \, \pi^2}  \, {C_{ll}} \nonumber \\
	& 
	+ \dfrac{{G_F} \,  \left( -3 \, M_h^2 - 6 \, m_t^2 + 4 \, M_W^2 + 3 \, M_Z^2 \right) }{4 \, \sqrt{2} \, \pi^2} \, {C_{\varphi l}^{(3)}} - \dfrac{{G_F} \,  \left[ 13 \, M_h^2 + 3 \,  \left( 4 \, m_t^2 - 8 \, M_W^2 + M_Z^2 \right)  \right] }{32 \, \sqrt{2} \, \pi^2} \, {C_{\varphi D}} \nonumber   \\
	& 
	+ \dfrac{9 \, \sqrt{\alpha_s}\, {G_F} \, M_h^2 }{2 \, \sqrt{2} \, \pi^{ \frac{3}{2}}}  \, {C_{G}} 
	+  \dfrac{\sqrt{\alpha_s} \, (\sqrt{2} {G_F})^{\frac{1}{2}} \, {m_t} }{2 \, \sqrt{2}  \, \pi^{\frac{3}{2}} }  \, {C_{tG}} 
	\Bigg\} + \dfrac{{G_F} \, m_t^2}{2 \, \sqrt{2} \, \pi^2} \, C_{tG}^2 
	+ \dfrac{\sqrt{\alpha_s}}{8 \, \pi^{\frac{3}{2}}} \, {C_{tG}} \, {C_{tH}}
	\nonumber  \\
	& + \dfrac{\sqrt{\alpha_s} \, (\sqrt{2} {G_F})^{\frac{1}{2}} \, {m_t}}{2 \, \sqrt{2} \, \pi^{\frac{3}{2}}} \, {C_{\varphi l}^{(3)}} \, {C_{tG}}
	- \dfrac{\sqrt{\alpha_s} (\sqrt{2} {G_F})^{\frac{1}{2}} \, {m_t}}{4 \, \sqrt{2} \, \pi^{\frac{3}{2}}} \, {C_{ll}} \, {C_{tG}} 
		- \epsilon \, \delta C_{\varphi G}^8
	\, .	
\end{align}
\endgroup

\section{Feynman rules}
\label{app:feynman}

In this appendix, we collect all needed Feynman rules valid up to ${\cal{O}}({1 / \LNP^4})$.
We adopt the notation of Ref.~\cite{Dedes:2017zog}, but choose the WCs  to be real and symmetric (e.g. $C^{uG\star}_{f_2f_1} = C^{uG}_{f_2f_1} \sim \delta_{f_1f_2}$). 
The remaining Feynman rules are only needed to ${\cal{O}}({1 / \LNP^2})$ in our calculation and can be
found in 
Ref.~\cite{Dedes:2017zog}.
For compactness, we present Feynman rules without inserting  the field redefinitions of Eqs~\eqref{eq:field_shifts}.

\subsection{Quark-Higgs-gauge vertices}

\begin{minipage}{0.28\linewidth}
	\includegraphics[scale=.38]{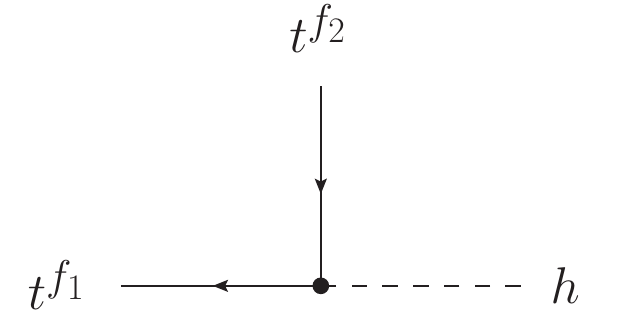}
\end{minipage}
\begin{minipage}{0.72\linewidth}
	\begin{flalign}
		\label{eq:app:feynman:tth}
		&  -  \frac{i}{\vev} \delta_{f_1 f_2} m_u R_\varphi^{-1} +\delta_{f_1 f_2} \frac{i \vev^2 \cht}{\sqrt{2}\Lambda^2} R_\varphi^{-1} &
	\end{flalign}
\end{minipage}

\subsection{Quark-gluon vertices}

\begin{minipage}{0.28\linewidth}
	\includegraphics[scale=.38]{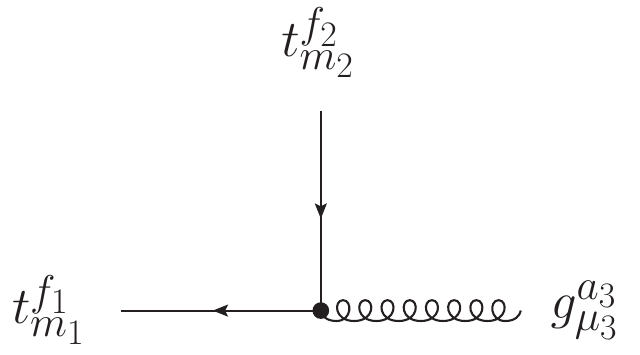}
\end{minipage}
\begin{minipage}{0.72\linewidth}
	\begin{flalign}
		& - i {\bar g}_s \delta_{f_1 f_2} {\cal T}_{m_1 m_2}^{a_3} \gamma^{\mu_3} - \sqrt2 \vev p_3^{\nu } {\cal T}_{m_1 m_2}^{a_3}  \sigma^{\mu_3 \nu } \frac{\cug}{\Lambda^2} R_g^{-1} &
	\end{flalign}
\end{minipage}

\vspace{0.3em}

\begin{minipage}{0.28\linewidth}
	\includegraphics[scale=.38]{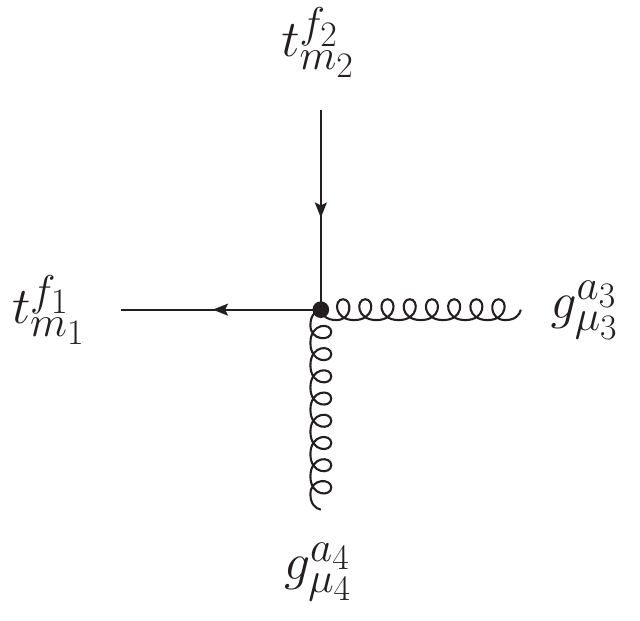}
\end{minipage}
\begin{minipage}{0.72\linewidth}
	\begin{flalign}
		\label{eq:feynman:uugg}
		& - i \sqrt2 \vev {\bar g}_s f_{a_3 a_4 {b_1}} {\cal T}_{m_1 m_2}^{{b_1}} \sigma^{\mu_3 \mu_4} \frac{\cug}{\Lambda^2} R_g^{-1}  &
	\end{flalign}
\end{minipage}

\vspace{0.3em}

\begin{minipage}{0.28\linewidth}
	\includegraphics[scale=.38]{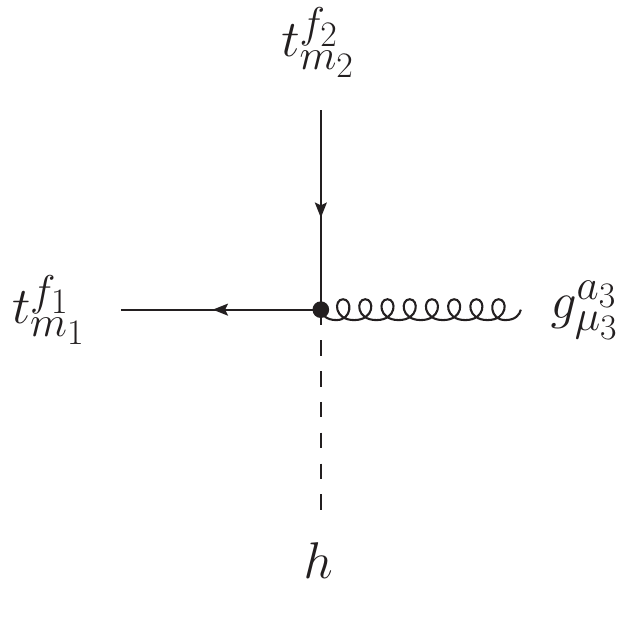}
\end{minipage}
\begin{minipage}{0.72\linewidth}
	\begin{flalign}
		&  - \sqrt2 p_3^{\nu } {\cal T}_{m_1 m_2}^{a_3}  \sigma^{\mu_3 \nu } \frac{\cug}{\Lambda^2} R_g^{-1} R_\varphi^{-1}&
	\end{flalign}
\end{minipage}

\subsection{Higgs-gauge vertices}

\begin{minipage}{0.28\linewidth}
	\includegraphics[scale=.38]{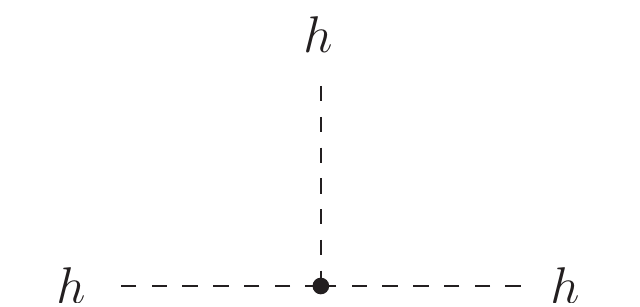}
\end{minipage}
\begin{minipage}{0.72\linewidth}
	\begin{flalign}
		&- 3 i \lambda \vev R_\varphi^{-3} + \frac{15 i \vev^3  \ch}{\Lambda^2}R_\varphi^{-3} - \frac{ i \vev  \chd}{\Lambda^2}   R_\varphi^{-3}  \big(  p_1\cdot p_2 + p_1 \cdot p_3  + p_2\cdot p_3 \big) & \nonumber \\ 
		&- \frac{ i \vev \chbox}{\Lambda^2} R_\varphi^{-3} \big( 3p_1^2 + 3 p_2^2 + 3 p_3^2 + 2 p_1\cdot p_2 + 2p_1 \cdot p_3  +2 p_2\cdot p_3\big) &
	\end{flalign}
\end{minipage}

\vspace{0.3em}

\begin{minipage}{0.28\linewidth}
	\includegraphics[scale=.38]{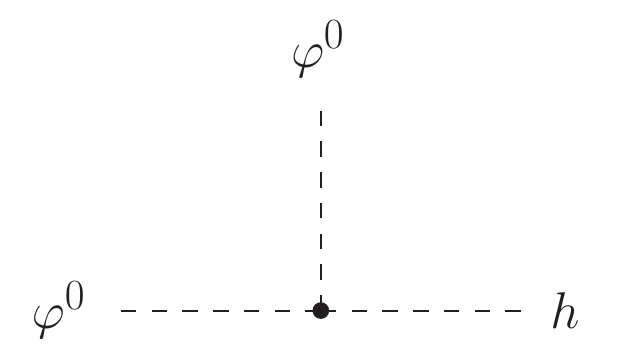}
\end{minipage}
\begin{minipage}{0.72\linewidth}
	\begin{flalign}
		&- i \lambda \vev R_\varphi^{-1} R_{\varphi^0}^{-2} +  \frac{3 i \vev^3  \ch}{\Lambda^2}  R_\varphi^{-1} R_{\varphi^0}^{-2} - \frac{i \vev \chbox}{\Lambda^2} R_\varphi^{-1} R_{\varphi^0}^{-2} \left(p_1^2 + p_2^2 + p_3^2 + 2 p_1 \cdot p_2 \right)  \nonumber  & \\
		&-  \frac{i \vev \chd }{\Lambda^2}  R_\varphi^{-1} R_{\varphi^0}^{-2} \left( p_1 \cdot p_2\right) & 
	\end{flalign}
\end{minipage}

\vspace{0.3em}

\begin{minipage}{0.28\linewidth}
	\includegraphics[scale=.38]{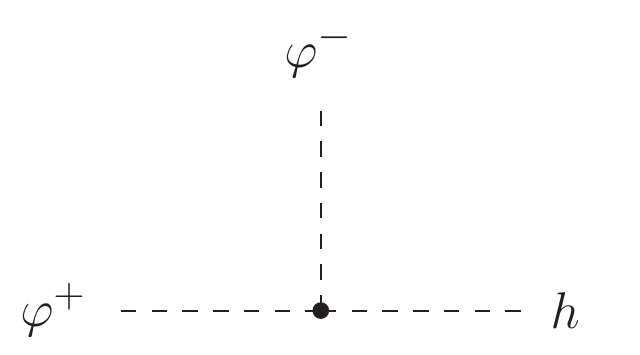}
\end{minipage}
\begin{minipage}{0.72\linewidth}
	\begin{flalign}
		& - i \lambda \vev R_\varphi^{-1} +   \frac{ 3 i \vev^3  \ch}{\Lambda^2} R_\varphi^{-1} -  \frac{i \vev \chbox}{\Lambda^2} R_\varphi^{-1} \left(p_1\nobreak\cdot\nobreak{}p_1 + 2 p_1\nobreak\cdot\nobreak{}p_2 + p_2\nobreak\cdot\nobreak{}p_2 + p_3\nobreak\cdot\nobreak{}p_3 \right) \nonumber & \\
		&- \frac{i \vev \chd}{2\Lambda^2} R_\varphi^{-1} (p_1\nobreak\cdot\nobreak{}p_3 + p_2\nobreak\cdot\nobreak{}p_3)   &
	\end{flalign}
\end{minipage}

\subsection{Higgs-gluon vertices}

\begin{minipage}{0.28\linewidth}
	\includegraphics[scale=.38]{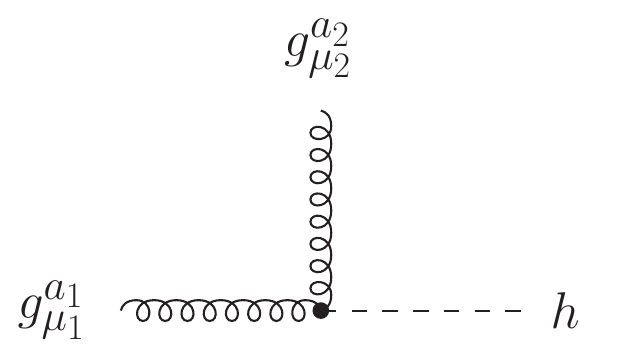}
\end{minipage}
\begin{minipage}{0.72\linewidth}
	\begin{flalign}
		&  + 4 i \vev \delta_{a_1 a_2} \frac{\chg}{\Lambda^2} R_g^{-2} R_\varphi^{-1} \Big(p_1^{\mu_2} p_2^{\mu_1} - (p_1 \cdot p_2) g^{\mu_1 \mu_2}\Big)  &
	\end{flalign}
\end{minipage}

\vspace{0.3em}

\begin{minipage}{0.28\linewidth}
	\includegraphics[scale=.38]{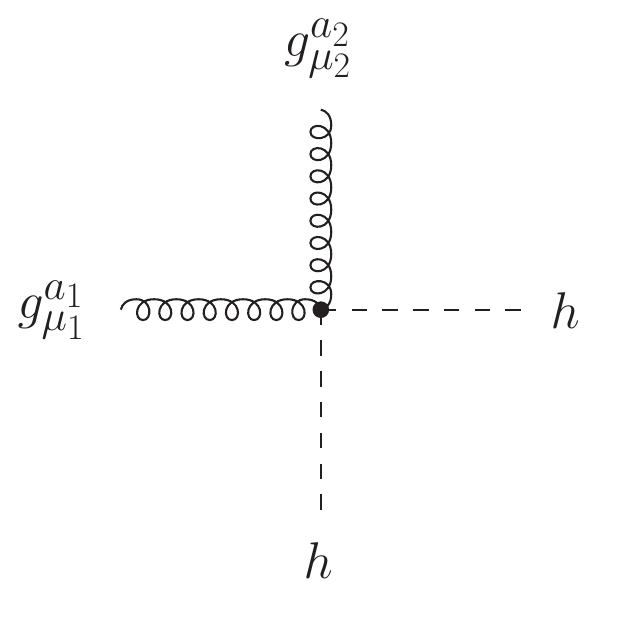}
\end{minipage}
\begin{minipage}{0.72\linewidth}
	\begin{flalign}
		&  + 4 i \delta_{a_1 a_2} \frac{\chg}{\Lambda^2}  R_g^{-2} R_\varphi^{-2} \Big(p_1^{\mu_2} p_2^{\mu_1} - (p_1 \cdot p_2) g^{\mu_1 \mu_2}\Big)  &
	\end{flalign}
\end{minipage}

\vspace{0.3em}

\begin{minipage}{0.28\linewidth}
	\includegraphics[scale=.38]{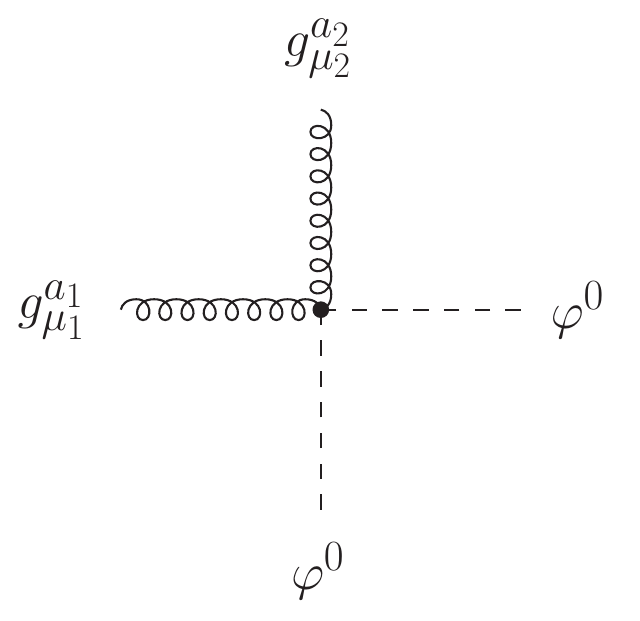}
\end{minipage}
\begin{minipage}{0.72\linewidth}
	\begin{flalign}
		&  + 4 i \delta_{a_1 a_2} \frac{\chg}{\Lambda^2}  R_g^{-2} R_{\varphi^0}^{-2} \Big(p_1^{\mu_2} p_2^{\mu_1} - (p_1 \cdot p_2) g^{\mu_1 \mu_2}\Big)  &
	\end{flalign}
\end{minipage}

\vspace{0.3em}

\begin{minipage}{0.28\linewidth}
	\includegraphics[scale=.38]{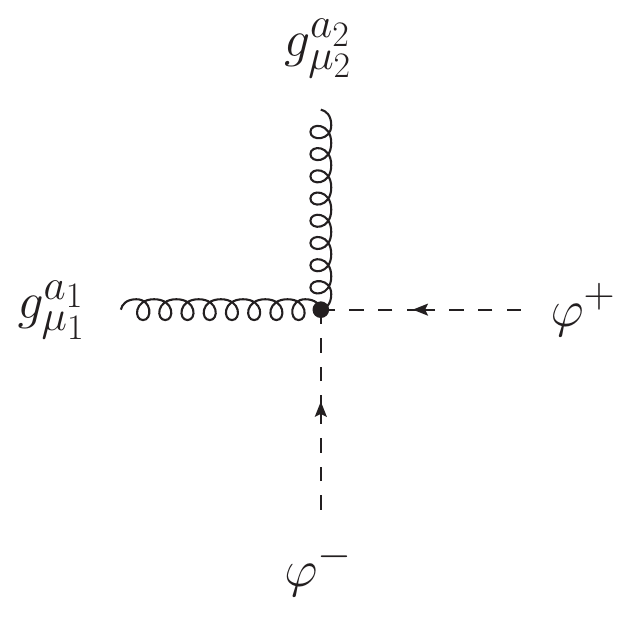}
\end{minipage}
\begin{minipage}{0.72\linewidth}
	\begin{flalign}
		&  + 4 i \delta_{a_1 a_2} \frac{\chg}{\Lambda^2}  R_g^{-2}  \Big(p_1^{\mu_2} p_2^{\mu_1} - (p_1 \cdot p_2) g^{\mu_1 \mu_2}\Big)  &
	\end{flalign}
\end{minipage}

\subsection{Gluon-gluon vertices}

\begin{minipage}{0.28\linewidth}
	\includegraphics[scale=.38]{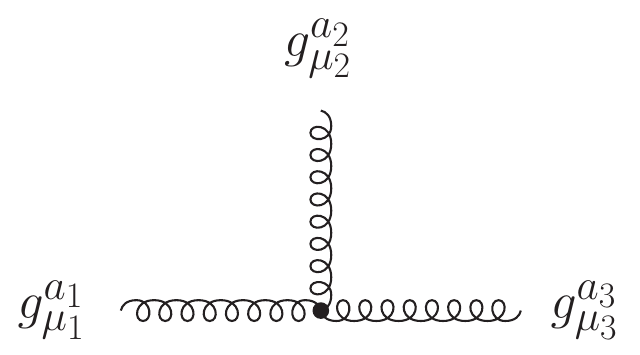}
\end{minipage}
\begin{minipage}{0.72\linewidth}
	\begin{flalign}
		&-\bar{g}_{s} f_{a_1 a_2 a_3} \left[\eta_{\mu_1 \mu_2}(p_{1}-p_{2})^{\mu_3} + \eta_{\mu_1 \mu_3}(p_{3}-p_{1})^{\mu_2} + \eta_{\mu_2 \mu_3}(p_{2}-p_{3})^{\mu_1}\right] \nonumber & \\
		&+\frac{6 \cg}{\Lambda^2} f_{a_1 a_2 a_3} R_g^{-3} \Big[p_{3}^{\mu_1}p_{1}^{\mu_2}p_{2}^{\mu_3}-p_{2}^{\mu_1}p_{3}^{\mu_2}p_{1}^{\mu_3} +\eta_{\mu_1 \mu_2}(p_1^{\mu_3}(p_2\cdot p_3)-p_2^{\mu_3}(p_1 \cdot p_3)) \nonumber & \\ 
		&\hphantom{=} +\eta_{\mu_2 \mu_3}(p_2^{\mu_1}(p_1\cdot p_3)-p_3^{\mu_1}(p_1\cdot p_2)) +\eta_{\mu_3 \mu_1}(p_3^{\mu_2}(p_1\cdot p_2)-p_1^{\mu_2}(p_2\cdot p_3))\Big]&
	\end{flalign}
\end{minipage}

\vspace{0.3em}

\begin{minipage}{0.28\linewidth}
	\includegraphics[scale=.38]{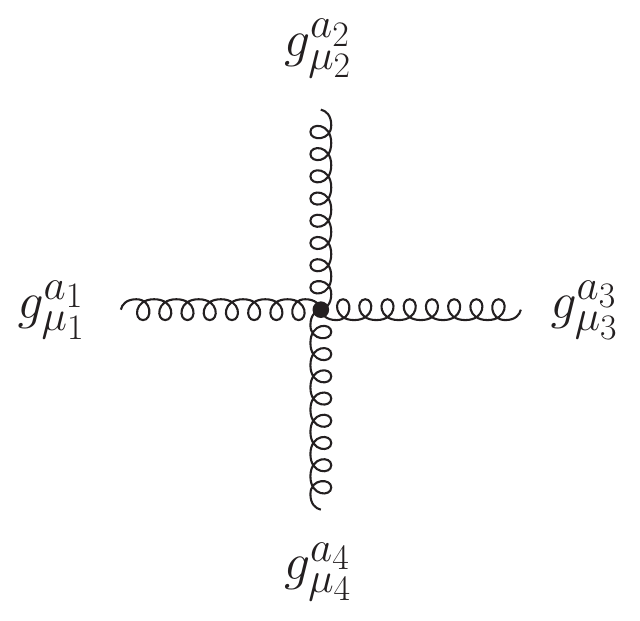}
\end{minipage}
\begin{minipage}{0.72\linewidth}
	\begin{flalign}
		&+ i {\bar g}_s^2 \Big(f_{a_1 a_2 {b_1}} f_{a_3
			a_4 {b_1}}\left(\eta_{\mu_1 \mu_4} \eta_{\mu_2 \mu_3} -
		\eta_{\mu_1 \mu_3} \eta_{\mu_2 \mu_4}\right) + f_{a_1 a_3 {b_1}} f_{a_2
			a_4 {b_1}} \left(\eta_{\mu_1 \mu_4} \eta_{\mu_2 \mu_3} \right.& \nonumber \\
		&\hphantom{==}- \left.
		\eta_{\mu_1 \mu_2} \eta_{\mu_3 \mu_4}\right) + f_{a_1 a_4 {b_1}} f_{a_2
			a_3 {b_1}} \left(\eta_{\mu_1 \mu_3} \eta_{\mu_2 \mu_4} -
		\eta_{\mu_1 \mu_2} \eta_{\mu_3 \mu_4}\right) \Big) \nonumber & \\[5pt]
		&-6i\bar{g}_{s} \frac{\cg}{\Lambda^2}  R_g^{-3} \Big( 
		f_{a_1 a_2 b_1} f_{a_3 a_4 b_1}[\eta_{\mu_1\mu_3}(p^{\mu_2}_{1}p^{\mu_4}_{2}+p^{\mu_2}_{4}p^{\mu_4}_{3}) +\eta_{\mu_2\mu_4}(p^{\mu_1}_{2}p^{\mu_3}_{1}+p^{\mu_1}_{3}p^{\mu_3}_{4}) \nonumber & \\
		&\hphantom{==} +\eta_{\mu_1\mu_2}(p^{\mu_3}_{2}p^{\mu_4}_{1}-p^{\mu_3}_{1}p^{\mu_4}_{2}) +\eta_{\mu_3\mu_4}(p^{\mu_1}_{4}p^{\mu_2}_{3}-p^{\mu_1}_{3}p^{\mu_2}_{4}) -\eta_{\mu_1\mu_4}(p^{\mu_2}_{1}p^{\mu_3}_{2}+p^{\mu_2}_{3}p^{\mu_3}_{4})\nonumber & \\
		&\hphantom{==}-\eta_{\mu_2\mu_3}(p^{\mu_1}_{2}p^{\mu_4}_{1}+p^{\mu_1}_{4}p^{\mu_4}_{3}) +(\eta_{\mu_1\mu_4}\eta_{\mu_2\mu_3}-\eta_{\mu_1\mu_3}\eta_{\mu_2\mu_4})(p_{1}\cdot p_{2}+p_{3} \cdot p_{4})] \nonumber & \\[5pt]
		&\hphantom{=}+f_{a_1 a_3 b_1} f_{a_2 a_4 b_1}[\eta_{\mu_1\mu_2}(p^{\mu_3}_{1}p^{\mu_4}_{3}+p^{\mu_3}_{4}p^{\mu_4}_{2})+\eta_{\mu_3\mu_4}(p^{\mu_1}_{3}p^{\mu_2}_{1}+p^{\mu_1}_{2}p^{\mu_2}_{4}) \nonumber & \\
		&\hphantom{==}+\eta_{\mu_1\mu_3}(p^{\mu_2}_{3}p^{\mu_4}_{1}-p^{\mu_2}_{1}p^{\mu_4}_{3})+\eta_{\mu_2\mu_4}(p^{\mu_1}_{4}p^{\mu_3}_{2}-p^{\mu_1}_{2}p^{\mu_3}_{4}) -\eta_{\mu_1\mu_4}(p^{\mu_2}_{3}p^{\mu_3}_{1}+p^{\mu_2}_{4}p^{\mu_3}_{2}) \nonumber & \\
		&\hphantom{==}-\eta_{\mu_2\mu_3}(p^{\mu_1}_{3}p^{\mu_4}_{1}+p^{\mu_1}_{4}p^{\mu_4}_{2}) +(\eta_{\mu_1\mu_4}\eta_{\mu_2\mu_3}-\eta_{\mu_1\mu_2}\eta_{\mu_3\mu_4})(p_{1}\cdot p_{3}+p_{2} \cdot p_{4})] \nonumber & \\[5pt]
		&\hphantom{=}+ f_{a_1 a_4 b_1} f_{a_2 a_3 b_1} [\eta_{\mu_1\mu_2}(p^{\mu_3}_{2}p^{\mu_4}_{3}+p^{\mu_3}_{4}p^{\mu_4}_{1})+\eta_{\mu_3\mu_4}(p^{\mu_1}_{2}p^{\mu_2}_{3}+p^{\mu_1}_{4}p^{\mu_2}_{1}) \nonumber & \\
		&\hphantom{==}+\eta_{\mu_1\mu_4}(p^{\mu_2}_{4}p^{\mu_3}_{1}-p^{\mu_2}_{1}p^{\mu_3}_{4})+\eta_{\mu_2\mu_3}(p^{\mu_1}_{3}p^{\mu_4}_{2}-p^{\mu_1}_{2}p^{\mu_4}_{3}) -\eta_{\mu_1\mu_3}(p^{\mu_2}_{4}p^{\mu_4}_{1}+p^{\mu_2}_{3}p^{\mu_4}_{2}) \nonumber & \\
		&\hphantom{==}-\eta_{\mu_2\mu_4}(p^{\mu_1}_{4}p^{\mu_3}_{1}+p^{\mu_1}_{3}p^{\mu_3}_{2}) +(\eta_{\mu_1\mu_3}\eta_{\mu_2\mu_4}-\eta_{\mu_1\mu_2}\eta_{\mu_3\mu_4})(p_{1}\cdot p_{4}+p_{2} \cdot p_{3}) ]	\Big) \hspace{-10pt} &
	\end{flalign}
\end{minipage}
\vskip 0.5in

\twocolumngrid
\bibliography{ggh_double}{}

\end{document}